# Quantum Fusion of Independent Networks Based on Multi-user Entanglement Swapping

Yiwen Huang[1,*], Yilin Yang[1,*], Hao Li[1,*], Jing Qiu[1], Zhantong Qi[1], Jiayu Wang[1], Yuting Zhang[1], Yuanhua Li[1,2], Yuanlin Zheng[1,3], and Xianfeng Chen[1,3,4,†]

[1]*State Key Laboratory of Advanced Optical Communication Systems and Networks, School of Physics and Astronomy, Shanghai Jiao Tong University, Shanghai 200240, China*

[2]*Department of Physics, Shanghai Key Laboratory of Materials Protection and Advanced Materials in Electric Power, Shanghai University of Electric Power, Shanghai 200090, China*

[3]*Shanghai Research Center for Quantum Sciences, Shanghai 201315, China*

[4]*Collaborative Innovation Center of Light Manipulation and Applications, Shandong Normal University, Jinan 250358, China*

[*]*These authors contribute equally to this work*

*Corresponding authors*: *Xianfeng Chen,* [†]*xfchen@sjtu.edu.cn*

## Abstract

With the advance development in quantum science[1-6], constructing a large-scale quantum network has become a hot area of future quantum information technology[7-11]. Future quantum networks promise to enable many fantastic applications and will unlock fundamentally new technologies in information security and large-scale computation. The future quantum internet is required to connect quantum information processors to achieve unparalleled capabilities in secret communication and enable quantum communication between any two points on Earth[12-14]. However, the existing quantum networks are basically constructed to realize the communication between the end users in their own networks. How to bridge different independent networks to form a fully-connected quantum internet becomes a pressing challenge for future networks. Here, we demonstrate the quantum fusion of two independent networks for the first time based on multi-user entanglement swapping, to merge two 10-user networks into a larger network with 18 users in quantum correlation layer. By performing the Bell state measurement between two non-neighboring nodes, the users from different networks can establish entanglement and ultimately every pair of the 18 users are able to communicate with each other using the swapped states. Our approach opens attractive opportunities for the establishment of quantum entanglement between remote nodes in different networks, which facilitates versatile quantum information interconnects and has great application in constructing large-scale intercity quantum communication networks.

## Introduction

As the paradigmatic quantum mechanical platform, entanglement-based quantum networks enable many dramatic applications such as secure communication[15,16], distributed quantum sensing[17,18] and fundamental tests of quantum mechanics[19-21]. Recently, the fully-connected quantum communication network based on quantum entanglement has garnered substantial interest because this type of network configuration enables multi users' communication with each other simultaneously when minimizing the infrastructure and hardware[22-25]. It is an important candidate for establishing a fully-connected quantum Internet, which may revolutionize the way of information exchange in the future. So far, the existing fully-connected networks constructed using dense wavelength division multiplexing (DWDM) technique are basically constructed to realize the communication between the end users in their own networks. To enable the communication between the end users in different independent networks in future, an efficient and feasible technique is required for quantum processors to achieve the quantum fusion of two independent networks, which means the two networks are merged into a larger network in the quantum correlation layer.

Entanglement swapping[26-28], which makes two independent quantum entangled states become entangled without direct interaction, provides a promising technology for bridging two independent quantum networks[29-31]. The entanglement swapping between two end nodes in different networks has been developed for connecting two different networks in the form of point-to-point topology, enabling two specific nodes' communication after the Bell state measurement (BSM)[32,33]. However, the network fusion based on the entanglement swapping of multi-user entangled states, which enables the communication among all the users in two different networks simultaneously, has not been demonstrated yet. Implementing the coherent fusion of two independent multi-user networks is extremely important for developing the future quantum internet. On the other hand, in the existing fully-connected networks constructed using DWDM technique or beam splitters[22-25], the number of quantum correlations and communication links becomes increasingly complex and grows quadratically as the number of users increases, resulting in the performance degradation and complexity increase of the quantum systems. Implementing Hong–Ou–Mandel (HOM) interference, a prerequisite condition of entanglement swapping, between multiple users is very challenging for these networks because high-visibility HOM interference requires the photons to be indistinguishable in all degrees of freedom including photon wavelength[34]. A novel and feasible scheme is needed to over this challenge for constructing a large-scale quantum network.

In this work, we present the first experimental demonstration of deterministic network fusion of two independent fully-connected networks based on multi-user entanglement swapping. We first develop a novel scheme, i.e., active temporal and wavelength multiplexing (ATWM) scheme, to construct two 10-user fully-connected quantum networks with single wavelength channel for each end node. By sending one node to a third party for BSM, every pair of users in different networks can generate polarization entanglement after entanglement swapping and the two independent networks are ultimately merged into a larger fully-connected quantum network with 18 end nodes.

Our approach provides a crucial capability of quantum communication in different networks and is advantageous for constructing a large-scale quantum internet that enable all the users' communication with each other.

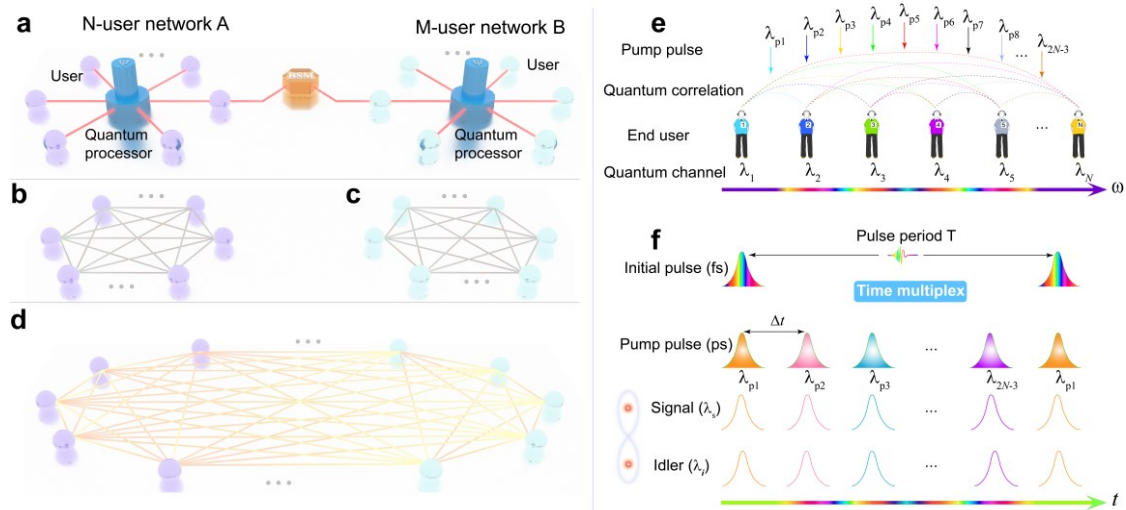

**Figure 1 Scheme of network fusion, network architecture and operation principle. a** Physical configuration of network fusion based on entanglement swapping. Two fully-connected networks send one node to the third-party for Bell state measurement. **b-c** Topological structures of networks A and B in quantum correlation layer, respectively. Both the networks have a fully-connected logical topology. **d** Topological structure of the new network after quantum fusion of networks A and B. **e** Operation principle of wavelength-multiplexing scheme for the unique design of pump pulses. **f** Operation principle of active temporal multiplexing scheme.

# Result
## Scheme of network fusion and network architecture
### a Network fusion based on entanglement swapping

To implement the fusion of two quantum networks, there are two technical challenges. First, two independent networks need to have a fully-connected topology so that their users can communicate with each other. Second, it requires high-quality entanglement swapping for all the involved entangled states simultaneously. Fig. 1 shows the over scheme of network fusion based on multi-user entanglement swapping, which is illustrated using three layers of abstraction. The Top layer shown in Fig.1a represents the physical topology of the fused network, mainly including three parts: physical components of network A, network B, and the third-party Charlie for BSM.

Fig. 1b-c illustrate the quantum correlation layers of network A and B with a fully-connected mesh before the network fusion, while Fig. 1d shows the overall quantum correlation layer of the integrated network after the network fusion. Two full-connected networks A and B, consisting of $M$ ($M \geq 2$) and $N$ ($N \geq 2$) end nodes, respectively, are merged into a larger full-connected network C with $M + N - 2$ end nodes as shown in Fig. 1d. In order to establish entanglement between the end nodes of two quantum network through entanglement swapping, two end nodes from two networks respectively are sent to the third-party Charil to perform a joint measurement and then the measured results are fed forward to the remaining end nodes of the two networks. By quantum entanglement swapping, two multi-user entangled networks consisting of $M$ and $N$ quantum nodes, respectively, can be merged into a new larger multi-user entangled network consisting of $M + N - 2$ quantum nodes.

**b Active temporal and wavelength multiplexing scheme**

To construct a $N$-user network with a fully connected graph in the quantum correlation, the quantum processor needs to prepare a minimum of $N(N-1)/2$ links to allocate them to the end users. Different from the pre-existing networks which allocate the entangled photon pairs to different users by using dense wavelength division multiplexers or beam splitters, we achieve the fully-connected topology of the network by directly pumping a nonlinear waveguide based on the ATWM scheme. The principle of our demonstration is illustrated in Fig. 1e-f. In our network, each user receives only one wavelength channel which consist of $N - 1$ temporal-separate photons entangled with other $N - 1$ user respectively. Assume the central wavelength of quantum channel for the user $N$ is $\lambda_N$. As shown in Fig. 1e, the quantum processor designs $2N - 3$ pump pulses with the central wavelength of $\lambda_{p1}$ to $\lambda_{p2N-3}$, respectively, where $\lambda_{p2N-3} = 1/(1/\lambda_{N-1} + 1/\lambda_N)$. Due to energy conservation during the spontaneous parametric down-conversion (SPDC) processes, the down-converted photon pairs are naturally frequency-time entangled and their spectrums are symmetric with respect to the central wavelength of their pump wavelengths. Thanks to the unique design of the pump wavelengths, all the symmetric wavelength channels for the end users can share a different entangled state with each other, creating the full-connected graph of the network. To effectively distinguish different photon pairs, we use the temporal multiplexing scheme to separate different pump pulses with a time interval of $\Delta t$, as shown in Fig.

1f. As a result, the down-converted photon pairs are also separated in time domain with the same time interval[35]. Then, one can distinguish different entangled states according to the photon arrival time, which can significantly improve the signal-to-noise ratio of the quantum systems. The most crucial technique in the ATWM scheme is that one need to ensure all the involved pump lasers can generate entangled photon pairs through the SPDC processes. We take advantage of the type-zero chirped quasi-phase-matching configuration and the high-efficiency property of the lithium niobate-on-insulator ridge waveguides to achieve this scheme.

## Experimental setup and network performance

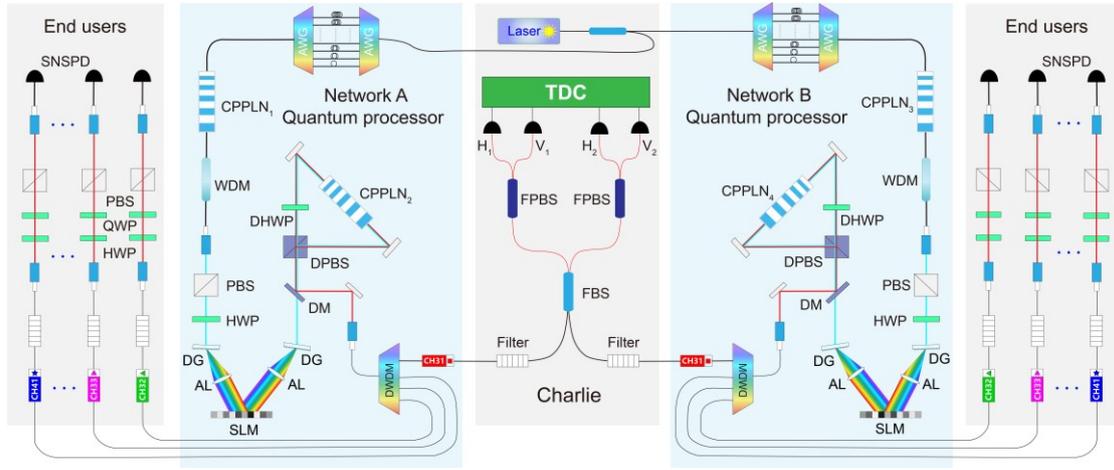

**Figure 2 Experimental setup for the network fusion of two fully-connected networks.** The quantum processors of networks A and B construct the polarization-entangled fully-connected networks by using the ATWM scheme. In each network, the femtosecond laser pulses from a mode-locked fiber laser are multiplexed into a series of pump pulses, and are frequency-doubled in a chirped periodically poled lithium niobate (CPPLN) waveguide. After passing through a phase compensator, the produced second harmonics create entangled photon pairs in the second CPPLN waveguide in a Sagnac loop. The quantum processor allocates the entangled photons to the end nodes using a ITU DWDM filter, and send one links to Charlie to performs the BSM with the combination of 50:50 fiber beam splitter (FBS), fiber polarization beam splitters (FPBS) and superconducting nanowire single photon detectors (SNSPD). TDC, time-to-digital converter; AWG, arrayed waveguide grating; WDM, wavelength division multiplexing filter; HWP, half wavelength plate; QWP, quarter wave plate; DG, diffraction grating; AL, achromatic lens; SLM, spatial light modulator; DM, dichroic mirror; DPBS, dual-wavelength PBS; GF, narrowband grating filter.

To demonstrate the versatility and flexibility of our approach, we perform the network fusion for two independent polarization entangled networks with 10 users of each. The experimental setup is depicted in Fig. 2. A femtosecond laser with a repetition frequency of 60 MHZ is first divided into two parts by a 50:50 fiber beam splitter (FBS) and then sent to the quantum processors A and B to construct the fully-connected networks using ATWM scheme. The second harmonics of pump lasers with specific design in wavelengths and time domain are injected into a CPPLN waveguide to generate polarization Einstein-Podolsky-Rosen (EPR) pairs in the Bell state $|\phi^+\rangle_n = 1/\sqrt{2}\,(|H\rangle_s|H\rangle_i + |V\rangle_s|V\rangle_i)$, via SPDC processes in a Sagnac loop, where $n$ is the wavelength number of the pump pulses, $s$ and $i$ denotes signal and idler photons. Finally, a thin-film DWDM filter is utilized to divide the entangled photons and allocate the unique ITU channels to the end users. The details of the experimental setup are provided in Methods. Thanks to the unique design of the pump lasers and high-efficiency of the CPPLN waveguides, each pair of quantum channels shares one pair of entangled photons with each other, forming a fully-connected mesh in quantum correlation layer of the network.

Before demonstrating the network fusion based on multi-user entanglement swapping, we first investigate the performance of the fully-connected network constructed using the proposed ATWM scheme. To show the advance of our scheme, we construct a twenty-user network as a demonstration. During this experiment, totally thirty-seven pump pulses with central wavelength varying from 1552.12 nm to 1537.79 nm are used to generate the correlative photon pairs via the SPDC process in the $CPPLN_2$ waveguide. A DWDM filter with ITU channels of CH31 to CH50 are leveraged to allocate the down-converted photons to the end users, respectively. To characterize the performance of the constructed network, we measure the coincidence counts between the twenty users after the entanglement distribution. The measurements are performed simultaneously for every ten users due to the limited detection equipment and the down-converted mean photon number for each photon pair is controlled to be $\mu \approx 0.01$ per pulse. The experimental results are shown in Fig. 3a, in which any two users have coincidence events with a count rate over $10^4$ in five seconds. It means that any two users share one pair of entangled photons with each other, indicating a fully-connected topological graph has been established. Due to the ATWM scheme, the end users can easily identify the entangled photons generated during each SPDC process and are able to calculate the coincidence count according to the photon arrival time. Fig. 3b shows the all temporal cross-correlation functions among the twenty end users. The cross-correlation peaks in the column of channel 31 ($N$) are the coincidence counts with channel 32 ($N$-1) to 50 under different delay time calculated by taking detection results of channel 31 ($N$) as the reference signal. These results confirm our proposed scheme can be used to construct a large-scale quantum network with a fully-connected topology structure. Our proposed fully-connected network can be encoded in time-bin (see Supplementary Information) and polarization degrees of freedom.

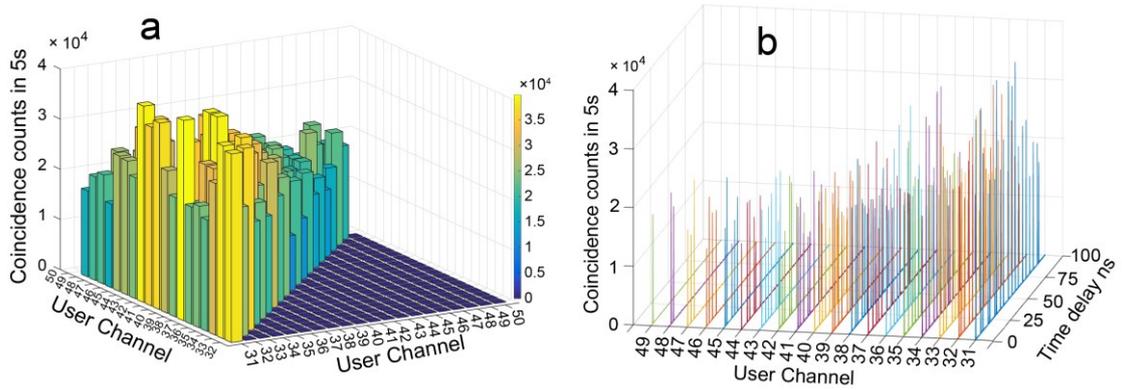

**Figure 3 Experimental results of the fully-connected network constructed using ATWM scheme. a** Measured two-photon coincidence counts between every pair of end users. **b** Temporal cross-correlations between the time traces of all the tween users. Each cross-correlation peak represents a pair of entangled photons shared by every pair of end users in the network.

To guarantee the high quality of the entanglement swapping, it is crucial for each network to ensure the end users share high-quality polarization entangled states with other users. The quantum processors first construct their ten-user polarization entangled networks using ATWM scheme as mentioned above. At this time, only seventeen pump pulses with the central wavelength varying from 1552.12 nm to 1545.7 nm are used to pump the CPPLN waveguides and the end users' channels turn into CH31 to CH40 for each network. To characterize the received polarization entangled states, each user in two networks is equipped with a SNSPD and a polarization state analyzer consisting of a HWP, a QWP and a PBS. We investigate the polarization entanglement of the networks by measuring the two-photon interference fringes in two mutually unbiased bases $|H\rangle/|V\rangle$ and $|D\rangle/|A\rangle$ by using the standard two-photon interference technique[36]. Fig. 4a and b show the typical interference fringes between CH31 and other users in networks A and B, respectively. The visibility of the interference fringes for each state is obtained to be higher than 95%, which exceeds the 70.7% local bound of the Bell's inequality[37] and reveals the existence of polarization entanglement in all available channel pairs. We calculate the fidelities of all the involved states in two networks as compared to ideal Bell state $|\phi^+\rangle = 1/\sqrt{2}\,(|H\rangle_s|H\rangle_i + |V\rangle_s|V\rangle_i)$ and obtain the fidelities to be greater than 96% for all the measured channel pairs in two networks, laying the foundations for high-quality network fusion based on entanglement swapping.

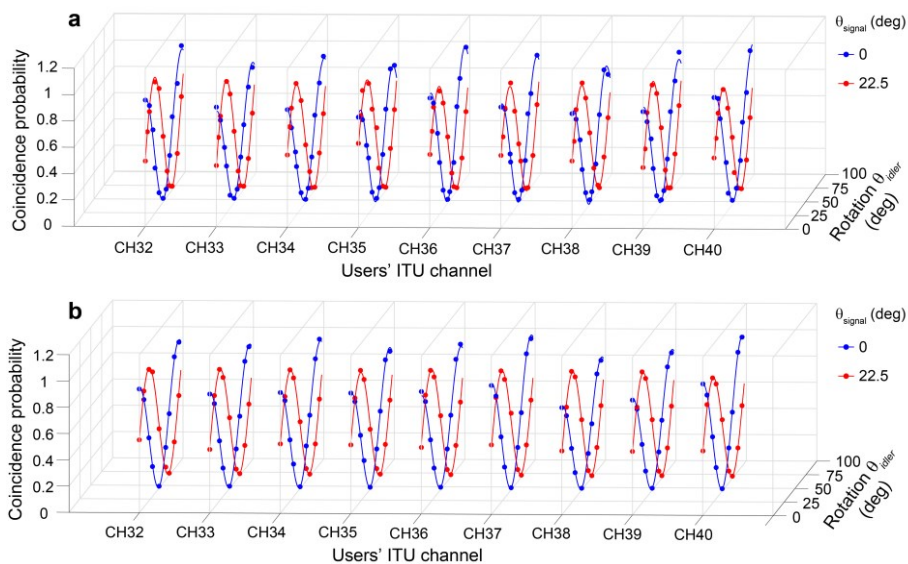

**Figure 4 Experimental two-photon interference for polarization entanglement between channel 31 and other users in the two networks. a** Two-photon coincidence count as a function of signal polarization under two nonorthogonal projection bases for network A. **b** Two-photon coincidence count as a function of signal polarization under two nonorthogonal projection bases for network B.

## Realization of multi-user Entanglement swapping

Next, we demonstrate the quantum fusion of the two polarization-entangled fully-connected networks by performing the multi-user entanglement swapping. To merge the two 10-user networks into an 18-user network in quantum correlation layer, both the quantum processors send the channel CH31 to the third-party Charlie to perform the BSM by interfering them on a FBS, as shown in Fig. 2. To enable the communication between every pair of the users in two independent networks, one need to establish entanglement between all the end nodes that never interacted by entanglement swapping. This requires a high-visibility HOM interference between all the involved photons corresponding to each quantum state in CH31 from the two networks. Thanks to the

ATWM scheme, each end node contains only one wavelength channel and every quantum state can be distinguished according to the photon arrival time, providing a prerequisite for high-visibility HOM interference among multiple users simultaneously. The two network providers can achieve the entanglement swapping between any two users from different networks by carefully adjusting the delay of pump pulses to overlap the corresponding CH31 signals on the FBS.

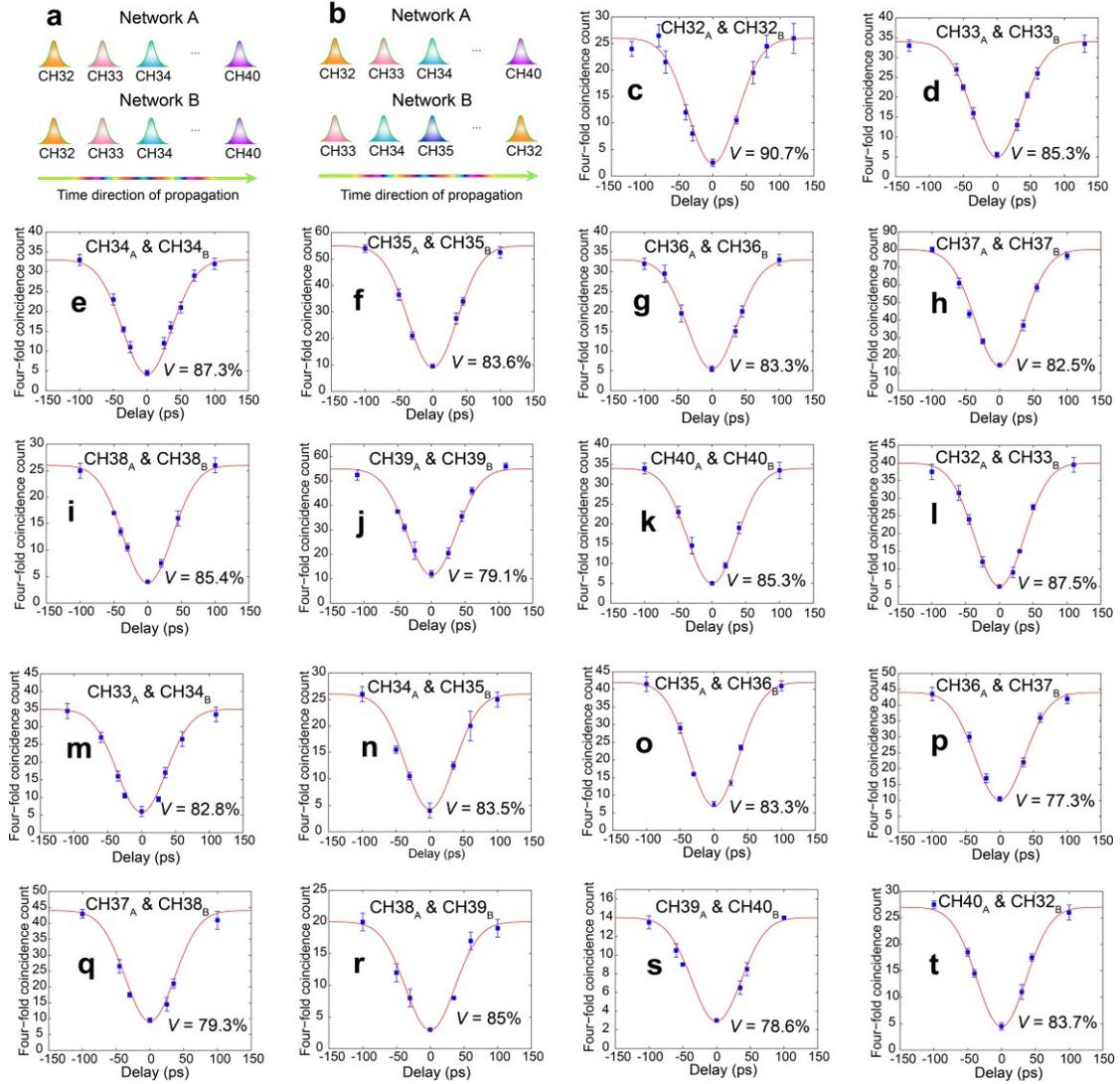

**Figure 5 Experimental HOM interference under two different delays. a-b** Temporal distribution of the channel 31 photons correlated with different channels from two networks. The channels in the same vertical column means that the channel 31 photons correlated with these channels overlap on the FBS, which are achieved by controlling the temporal distribution of the pump pulses in the quantum processors. **c-k** HOM interference for different photon pairs under the temporal distribution of the channel 31 photons shown in **a**. **l-t** HOM interference for different

photon pairs under the temporal distribution of the channel 31 photons shown in **b**. Each four-fold coincidence count is measured for 400 s and the error bars indicate one standard deviation.

As a reliability check before performing the experimental test of bilocal hidden variable models for the swapped entangled pairs, we first measure the HOM interference for the two networks. To achieve a high-visibility interference between two networks, we exploit two kinds of narrowband grating filters with a bandwidth of 0.1 nm and 0.2 nm for CH31 photons and other channels, respectively, to suppress the spectral distinguishability. See the Supplemental Material for further details about the experimental settings. The measured HOM interferences are shown in Fig. 5, in which Fig .5c-k and Fig. 5l-t are the HOM interferences for the time settings in Fig. 5a and Fig.5b, respectively. The visibilities defined as $V_{dip} = 1 - R_0/R_t$, where $R_0$ and $R_t$ are fitted counts at zero and infinite delays, respectively, are obtained to be over 75% for all the entangled state with the highest and average visibilities of 90.7% and $(83.5 \pm 3.4)\%$, respectively, which significantly break the classical boundary of 50%.

## Characterization of the entangled states

We confirm successful entanglement swapping of two networks by testing the entanglement of the previously uncorrelated photons for all the involved photon pairs. The specific states of the swapped entangled pairs depend on the result of the BSM. Consider the successful generation of both initial entangled states $|\phi^+\rangle_n = 1/\sqrt{2}\,(|H\rangle_{CH31_A}|H\rangle_{\lambda_n} + |V\rangle_{CH31_A}|V\rangle_{\lambda_n})$ and $|\phi^+\rangle_m = 1/\sqrt{2}\,(|H\rangle_{CH31_B}|H\rangle_{\lambda_m} + |V\rangle_{CH31_B}|V\rangle_{\lambda_m})$, where $CH31_A$ and $CH31_B$ represent the channel 31 photons from network A and B, respectively, and $\lambda_n$ and $\lambda_m$ represent the correlated idler photons in channel 32 to 40 of networks A and B. When overlapping the photons $CH31_A$ and $CH31_B$ on the FBS, the overall states of the entangled photon pairs would be cast into:

$$|\psi\rangle = |\phi^+\rangle_n \otimes |\phi^+\rangle_m$$
$$= \frac{1}{2}(|H\rangle_{CH31_A}|H\rangle_{\lambda_n} + |V\rangle_{CH31_A}|V\rangle_{\lambda_n}) \otimes (|H\rangle_{CH31_B}|H\rangle_{\lambda_m} + |V\rangle_{CH31_B}|V\rangle_{\lambda_m})$$
$$= \frac{1}{2}(|\phi^+\rangle_{CH31_A CH31_B}|\phi^+\rangle_{\lambda_n \lambda_m} + |\phi^-\rangle_{CH31_A CH31_B}|\phi^-\rangle_{\lambda_n \lambda_m} + |\psi^+\rangle_{CH31_A CH31_B}|\psi^+\rangle_{\lambda_n \lambda_m}$$
$$+ |\psi^-\rangle_{CH31_A CH31_B}|\psi^-\rangle_{\lambda_n \lambda_m}). \tag{1}$$

Charlie implements the Bell-state measurement by HOM interference at a 50:50 FBS and subsequent two FPBS, as shown in Fig. 2. With these experimental configurations, one can discriminate two of the four Bell states $|\psi^+\rangle_{CH31_A CH31_B} = 1/\sqrt{2}\,(|H\rangle_{CH31_A}|V\rangle_{CH31_B} + |V\rangle_{CH31_A}|H\rangle_{CH31_B})$ and $|\psi^-\rangle_{CH31_A CH31_B} = 1/\sqrt{2}\,(|H\rangle_{CH31_A}|V\rangle_{CH31_B} - |V\rangle_{CH31_A}|H\rangle_{CH31_B})$, which is the optimum efficiency possible with linear optics. A twofold coincidence detection event between either $D_{H_1}$ and $D_{V_2}$ or $D_{V_1}$ and $D_{H_2}$ indicates a projection on $|\psi^-\rangle_{CH31_A CH31_B}$, while a coincidence detection event between either $D_{H_1}$ and $D_{V_1}$ or $D_{H_2}$ and $D_{V_2}$ indicates a projection on $|\psi^+\rangle_{CH31_A CH31_B}$. It is worth noting that the BSM results for different photon pairs can be easily identify according to the photon arrival time owing to the ATWM scheme. We select the Bell state $|\psi^-\rangle_{CH31_A CH31_B}$ as an example to demonstrate the multi-user entanglement swapping between two networks. Measuring two-photon interference under two mutually unbiased bases can detect whether the quantum state has breaking inequalities. We control the projection basis of idle photons in network A as $|H\rangle/|V\rangle$ and $|D\rangle/|A\rangle$, respectively, and measure the statistical quadruple coincidence counts by scanning the half-wave plate angle of idle photons in network B. The typical experimental results are shown in the Fig. 6. We obtain the average visibilities of the interference fringes to be $V_{swapped} = (R_{max} - R_{min})/(R_{max} + R_{min}) = (81.2 \pm 2.0)\%$ and $(79.8 \pm 5.5)\%$ for the swapped entangled pairs CH32$_A$ & CH32$_B$ and CH32$_A$ & CH33$_B$, respectively, for the temporal settings in Fig. 5a and Fig. 5b. Such a visibility achieved in our experiment clearly exceeds the classical bound of CHSH inequality and reveal the polarization entanglement between the swapped entangled pairs. Moreover, we calculate the fidelities of the entangled states after the entanglement swapping using $F = (3V_{swapped} + 1)/4$. The typical results are shown in Fig. 6c, in which one can obtain an average fidelity of $(84.5 \pm 2)\%$. These results prove that polarization entanglement can be generated between the previously uncorrelated users in two networks by controlling the delay of pump pulses to overlap the signal photons of different photon pairs on the FBS. Ultimately, the two 10-user fully-connected networks are merged into an 18-user network in the quantum correlation layer after the multi-user entanglement swapping. All the end users can communicate with each other by using the entanglement-based quantum key distribution protocol. See Supplementary Information for the analysis of quantum key distribution between every pair of users in the fused 18-user network.

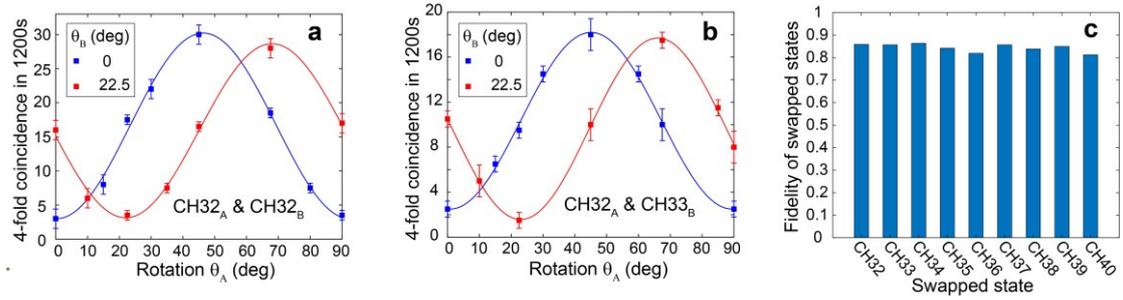

**Figure 6 Experimental two-photon interference for polarization entanglement after multi-user entanglement swapping. a** Four-fold coincidence counts as a function of signal polarization for the swapped photon pair CH32$_A$ & CH32$_B$. **b** Four-fold coincidence counts as a function of signal polarization for the swapped photon pair CH32$_A$ & CH33$_B$. **c** Measured fidelities of the swapped states. CH32 to CH40 in the x-axis represent the swapped states of CH32$_A$ & CH32$_B$ to CH40$_A$ & CH40$_B$, respectively. The error bars indicate one standard deviation.

## Discussion

Here we highlight the substantial advantages of our approach. First, our proposed fully-connected quantum networks based on ATWM scheme have a distinct advantage in scalability and signal-to-noise ratio over the existing networks constructed by using DWDM technique or beam splitters. To add a new user into the latter, the quantum processor needs to multiplex a large number of quantum channels into each user's fiber, which means a considerable change to the quantum system. In our scheme, only two more wavelength lasers are required to multiplex into the pump pulses for the SPDC processes thanks to the photon energy conservation. Besides, owing to the ingenious design of the pump pulses in temporal domain, the SPDC photons generated by different pump pulses can be detected in different photon time slots and the end users can clearly distinguish the photon detection results of each entangled state according to the photon arrival time. This characteristic would significantly improve the signal-to-noise ratio and enable the end users' two-party communication. Moreover, our proposed network has the nearly lowest system loss and most compact system architecture because only one DWDM filter is need to allocate the entangled photons to the end users. These salient characteristics make the ATWM scheme more suitable to construct a large-scale fully-connected network. Second, each user contains only one wavelength channel, which is conducive to subsequent quantum operations, such as quantum interference and frequency-up conversion for quantum storage. Last but most importantly, our proposed network fusion scheme based on multi-user entanglement swapping enables the connection of the remote users in different networks, merging two independent networks into a larger network for the first time. This approach can realize mutual communication between all users in different networks, which has important application prospects in constructing large-scale quantum networks in future.

**Conclusion**

In summary, we have successfully realized the quantum fusion of two independent 10-user networks based on multi-user entanglement swapping, merging them into a larger network with 18 users in quantum correlation layer. The results show that the proposed ATWM scheme can be used to construct high-performance larger-scale networks with a fully-connected topological structure. By performing the BSM between two end nodes, the users from different networks can establish entanglement after entanglement swapping, and ultimately every pair of all the 18 users can communicate with each other. Our approach opens attractive opportunities for the establishment of quantum entanglement between remote nodes in different networks, which facilitates versatile quantum information interconnects and has great application in constructing large-scale intercity quantum communication networks.

**Materials and methods**

**Multi-wavelength polarization-entangled photon-pair source.** To construct polarization entangled fully-connected networks using ATWM scheme, we develop a special polarization entangled source based on a polarization compensator and a Sagnac loop, as shown in Fig. 2, to generate the EPR Bell states $|\phi^+\rangle = 1/\sqrt{2} \left( |H\rangle_s |H\rangle_i + |V\rangle_s |V\rangle_i \right)$ for all the pump pulses. In the physical topology layer of each quantum processor, the femtosecond laser pulses are multiplexed into a series of pump pulses with the wavelength and time intervals of $\Delta \lambda_p = 0.4 \ nm$ and $\Delta t \approx 300 \ ps$, respectively, by using arrayed waveguide gratings whose central wavelengths are identified by International Telecommunication Union (ITU). Then, the pump pulses are frequency doubled in a chirped periodically poled lithium niobate (CPPLN) waveguide by second-harmonic generation. The CPPLN waveguides are fabricated by using the method of UV lithography and deep dry etching[38] and enable effective SH generation with over 15-nm pump wavelength range (See the Supplemental Material). The remanent pump pulses are suppressed by a wavelength division multiplexing filter with an extinction ratio of 180 dB. The multi-wavelength SH lasers first go through a Fourier-transform setup which consists of two diffraction gratings, two achromatic lens and a spatial light modulator (SLM) to compensate the polarization and phase of different wavelength pulses. The photons with diagonal polarization are first spectrally dispersed by a grating and then focused to an elongated spot with the wavelength varying from $\lambda_{min}$ to $\lambda_{max}$ by an achromatic lens. A horizontal SLM with a resolution of 1920 × 1200 pixels is placed in the focal plane of the lens to implement the phase control between the horizontal and vertical components for different wavelength channels. One can modulate the polarization of the SHG pulses at different wavelengths by adjusting the corresponding phase of the diagram loaded on the SLM according to the spatial position of the spot. Then, the SH lasers are used to pump a CPPLN waveguide placed inside a polarization Sagnac loop to generate the polarization entangled photon pairs. A dichroic mirror is leveraged to separate the photon pairs from the SHG pulses and the entangled photon pairs are ultimately allocated to the end users by using a ITU DWDM filter. See the Supplemental Material for further details about the central wavelength of the pump lasers and quantum channels. The photons distributed to all the end users are detected by superconducting nanowire single-photon detectors (SNSPD) with the detect efficiency varying from 60% to 80% and dark count rate of 40 to 100 per second. The photon detection results are recorded by a time-to-digital converter (TDC) which is synchronized by the electric signal from the mode-locked fiber laser.

## Acknowledgements

This work is supported in part by the National Natural Science Foundation of China (Grant Nos. 12192252 and 12074155), The Foundation for Shanghai Municipal Science and Technology Major Project (Grant No. 2019SHZDZX01-ZX06).

## Conflict of interest

The authors declare no competing interests.

## Data Availability

The data that support the findings of this study are available from the corresponding author upon reasonable request.

## Code Availability

The code used for modelling the data is available from the corresponding authors on reasonable request

## Contributions

X.C. led the project since its conception and supervised all experiments. Y.H., Y.Y. and Z.Q. performed the experiment and data analysis. H.L., J.Q., J.W. and Y.Z. develop the CPPLN waveguides. All authors participated in discussions of the results. Y.H. prepared the manuscript with assistance from all other co-authors. Y.L., Y.Z. and X.C. provided revisions.

## References

1. F. Xu, X. Ma, Q. Zhang, H. K., Lo, J. W. Pan, Secure quantum key distribution with realistic devices. *Rev. Mod. Phys.* **92**, 025002 (2020).
2. W. Li, L. Zhang, H. Tan, Y. Lu, S.-K. Liao, J. Huang, H. Li, Z. Wang, H.-K. Mao, B. Yan, Q. Li, Y. Liu, Q. Zhang, C.-Z. Peng, L. You, F. Xu, J.-W. Pan, High-rate quantum key distribution exceeding 110 Mb s$^{-1}$. *Nat. Photonics* **17**, 416-421 (2023).
3. F. Grünenfelder, A. Boaron, G. V. Resta, M. Perrenoud, D. Rusca, C. Barreiro, R. Houlmann, R. Sax, L. Stasi, S. El-Khoury, E. Hänggi, N. Bosshard, F. Bussières, H. Zbinden, Fast single-photon detectors and real-time key distillation enable high secret-key-rate quantum key distribution systems. *Nat. Photonics* **17**, 422-426 (2023).
4. P. C. Humphreys, N. Kalb, J. P. Morits, R. N. Schouten, R. F. Vermeulen, D. J. Twitchen, M. Markham, R. Hanson, Deterministic delivery of remote entanglement on a quantum network. *Nature* **558**, 268-273 (2018).
5. S. Wang, Z. Q. Yin, D. Y. He, W. Chen, R. Q. Wang, P. Ye, Y. Zhou, G.-J. Fan-Yuan, F.-X.Wang, W. Chen, Y.-G. Zhu, P. V. Morozov, A. V. Divochiy, Z. Zhou, G.-C. Guo, Z. F. Han, Twin-field quantum key distribution over 830-km fibre. *Nat. Photonics* **16**, 154-161 (2022).
6. Y. Zhong, H.-S. Chang, A. Bienfait, É. Dumur, M.-H. Chou, C. R. Conner, J. Grebel, R. G. Povey, H. Yan, D. I. Schuster, A. N. Cleland, Deterministic multi-qubit entanglement in a quantum network. *Nature* **590**, 571-575 (2021).
7. A. Kržič, S. Sharma, C. Spiess, U. Chandrashekara, S. Töpfer, G. Sauer, L. J. González-Martín del Campo, T. Kopf, S. Petscharnig, T. Grafenauer, R. Lieger, B. Ömer, C. Pacher, R. Berlich, T.Peschel, C. Damm, S. Risse, M. Goy, D. Rieländer, A. Tünnermann, F. Steinlechner, Towards metropolitan free-space quantum networks. *NPJ Quantum Inf.* **9**, 95 (2023).
8. D. Ribezzo, M. Zahidy, I. Vagniluca, N. Biagi, S. Francesconi, T. Occhipinti, L. K. Oxenløwe, M. Lončarić, I. Cvitić, M. Stipčević, Ž. Pušavec, R. Kaltenbaek, A. Ramšak, F. Cesa, G. Giorgetti, F. Scazza, A. Bassi, P. D. Natale, F. S. Cataliotti, M. Inguscio, D. Bacco, A. Zavatta, Deploying an Inter-European Quantum Network. *Adv. Quantum Technol.* **6**, 2200061 (2023).
9. Y. A. Chen, et al. An integrated space-to-ground quantum communication network over 4,600

kilometres. *Nature* **589**, 214-219 (2021).

10. Y. Li, Y. Huang, T. Xiang, Y. Nie, M. Sang, L. Yuan, X. Chen, Multiuser time-energy entanglement swapping based on dense wavelength division multiplexed and sum-frequency generation. *Phys. Rev. Lett.* **123**, 250505 (2019).

11. M. Pompili, S. L. Hermans, S. Baier, H. K. Beukers, P. C. Humphreys, R. N. Schouten, R. F. L. Vermeulen, M. J. Tiggelman, L. Dos Santos Martins, B. Dirkse, S. Wehner, R. Hanson, Realization of a multinode quantum network of remote solid-state qubits. *Science* **372**, 259-264 (2021).

12. S. Wehner, D. Elkouss, R. Hanson, Quantum internet: A vision for the road ahead. *Science* **362**, eaam9288 (2018).

13. S.-H. Wei, B. Jing, X.-Y. Zhang, J.-Y. Liao, C.-Z. Yuan, B.-Y. Fan, C. Lyu, D.-L. Zhou, Y. Wang, G.-W. Deng, H.-Z. Song, D. Oblak, G.-C. Guo, Q. Zhou, Towards real-world quantum networks: a review. *Laser Photonics Rev.* **16**, 2100219 (2022).

14. C. Simon, Towards a global quantum network. *Nat. Photonics* **11**, 678-680 (2017).

15. S. L. N. Hermans, M. Pompili, H. K. C. Beukers, S. Baier, J. Borregaard, R. Hanson, Qubit teleportation between non-neighbouring nodes in a quantum network. *Nature* **605**, 663-668 (2022).

16. A. S. Cacciapuoti, M. Caleffi, F. Tafuri, F. S. Cataliotti, S. Gherardini, G. Bianchi, Quantum internet: Networking challenges in distributed quantum computing. *IEEE Network* **34**, 137-143 (2019).

17. C. L. Degen, F. Reinhard, P. Cappellaro, Quantum sensing. Rev. Mod. Phys. 89, 035002 (2017).

18. X. Guo, C. R. Breum, J. Borregaard, S. Izumi, M. V. Larsen, T. Gehring, M. Christandl, J. S. Neergaard-Nielsen, U. L. Andersen. Distributed quantum sensing in a continuous-variable entangled network. *Nat. Phys.* **16**, 281-284 (2020).

19. E. Polino, D. Poderini, G. Rodari, I. Agresti, A. Suprano, G. Carvacho, E. Wolfe, A. Canabarro, G. Moreno, G. Milani, R. W. Spekkens, R. Chaves, F. Sciarrino, Experimental nonclassicality in a causal network without assuming freedom of choice. *Nat. Commun.* **14**, 909 (2023).

20. N. N. Wang, A. Pozas-Kerstjens, C. Zhang, B. H. Liu, Y.F. Huang, C. F. Li, G. C. Guo, N. Gisin, A. Tavakoli, Certification of non-classicality in all links of a photonic star network without assuming quantum mechanics. *Nat. Commun.* **14**, 2153 (2023).

21. Q. C. Sun, Y. F. Jiang, B. Bai, W. Zhang, H. Li, X. Jiang, J. Zhang, L. You, X. Chen, Z. Wang, Q. Zhang, J. Fan, J. W. Pan, Experimental demonstration of non-bilocality with truly independent sources and strict locality constraints. *Nat. Photonics* **13**, 687-691 (2019).

22. S. Wengerowsky, S. K. Joshi, F. Steinlechner, H. Hübel, R. Ursin, An entanglement-based wavelength multiplexed quantum communication network. *Nature* **564,** 225-228 (2018).

23. S. K. Joshi, D. Aktas, S. Wengerowsky, M. Lončarić, S. P. Neumann, B. Liu, T. Scheidl, G. C. Lorenzo, Z. Samec, L. Kling, A. Qiu, M. Razavi, M. Stipcevic, J. G. Rarity, R. Ursin, A trusted node-free eight-user metropolitan quantum communication network. *Sci. Adv.* **6**, eaba0959 (2020).

24. X. Liu, X. Yao, R. Xue, H. Wang, H. Li, Z. Wang, L. You, X. Feng, F. Liu, K. Cui, Y. Huang, W. Zhang, An entanglement-based quantum network based on symmetric dispersive optics quantum key distribution. A*PL Photonics* **5**, 076104 (2020).

25. J. H. Kim, J. W. Chae, Y. C. Jeong, Y. H. Kim, Quantum communication with time-bin entanglement over a wavelength-multiplexed fiber network. *APL Photonics* **7**, 016106 (2022).

26. J. W. Pan, D. Bouwmeester, H. Weinfurter, A. Zeilinger, Experimental entanglement swapping: entangling photons that never interacted. *Phys. Rev. Lett.* **80**, 3891 (1998).

27. F. Samara, N. Maring, A. Martin, A. S. Raja, T. J. Kippenberg, H. Zbinden, R. Thew Entanglement swapping between independent and asynchronous integrated photon-pair sources. *Quantum Science and Technology* **6**, 045024 (2021).

28. S. Liu, Y. Lou, Y. Chen, J. Jing, All-optical entanglement swapping. *Phys. Rev. Lett.* **128**, 060503 (2022).

29. R. Kaltenbaek, R. Prevedel, M. Aspelmeyer, A. Zeilinger, High-fidelity entanglement swapping with fully independent sources. *Phys. Rev. A* **79**, 040302 (2009).

30. C. Y. Lu, T. Yang, J. W. Pan, Experimental multiparticle entanglement swapping for quantum networking. *Phys. Rev. Lett.* **103**, 020501 (2009).

31. E. Shchukin, P. van Loock, Optimal entanglement swapping in quantum repeaters. *Phys. Rev. Lett.* **128**, 150502 (2022).

32. G. Guccione, T. Darras, H. Le Jeannic, V. B. Verma, S. W. Nam, A. Cavaillès, J. Laurat, Connecting heterogeneous quantum networks by hybrid entanglement swapping. *Science advances* **6**, eaba4508 (2020).

33. Q. C. Sun, Y. F. Jiang, Y. L. Mao, L. X. You, W. Zhang, W. J. Zhang, X. Jiang, T. Y. Chen, H. Li, Y. D. Huang, X. F. Chen, Z. Wang, J. Fan, Q. Zhang, J. W. Pan, Entanglement swapping over 100 km optical fiber with independent entangled photon-pair sources. *Optica* **4**, 1214-1218 (2017).

34. C. K. Hong, Z. Y. Ou, L. Mandel, Measurement of subpicosecond time intervals between two photons by interference. *Phys. Rev. Lett.* **59**, 2044 (1987).

35. C. Xiong, X. Zhang, Z. Liu, M. J. Collins, A. Mahendra, L. G. Helt, M.J. Steel, D.-Y. Choi, C. J. Chae, B. J. Eggleton, Active temporal multiplexing of indistinguishable heralded single photons. *Nat. Commun.* **7**, 10853 (2016).

36. P. G. Kwiat, K. Mattle, H. Weinfurter, A. Zeilinger, A. V. Sergienko, Y. Shih, New high-intensity source of polarization-entangled photon pairs. *Phys. Rev. Lett.* **75**, 4337 (1995).

37. J. F. Clauser, M. A. Horne, A. Shimony, R. A. Holt, Proposed experiment to test local hidden-variable theories. *Phys. Rev. Lett.* **23**, 880 (1969).

38. Y. Zhang, H. Li, T. Ding, Y. Huang, L. Liang, X. Sun, Y. Tang, J. Wang, S. Liu, Y. Zheng, X. Chen, Scalable, fiber-compatible lithium-niobate-on-insulator micro-waveguides for efficient nonlinear photonics. *Optica* **10**, 688-693 (2023).